\documentclass[aps,prl,twocolumn,superscriptaddress,showpacs,floats]{revtex4-1}
\usepackage{graphicx}
\usepackage{amsmath,graphicx,dcolumn}
\usepackage{hyperref}
\usepackage{datetime}
\usepackage{color}
\usepackage{textcomp}
\usepackage{ulem}
\usepackage{epstopdf}

\newcommand*{\ur}{URu$_2$Si$_2$}

\begin{document}

\title{Exceptional Ising Magnetic Behavior of Itinerant Spin-polarized Carriers in  {\ur} }

\author{Miros{\l}aw Werwi{\'n}ski }
\affiliation{Department of Physics and Astronomy, Uppsala University, Box 516, S-75120 Uppsala, Sweden }

\author{Jan Rusz}
\affiliation{Department of Physics and Astronomy, Uppsala University, Box 516, S-75120 Uppsala, Sweden }

\author{John A. Mydosh}
\affiliation{Kamerlingh Onnes Laboratory, Leiden University, NL-2300 RA Leiden, The Netherlands }

\author{Peter M. Oppeneer}
\affiliation{Department of Physics and Astronomy, Uppsala University, Box 516, S-75120 Uppsala, Sweden }

\date{\today}

\begin{abstract}
We show on the basis of electronic structure calculations that
the uranium 5$f$ magnetic moment in {\ur} exhibits a unique Ising behavior, which surprisingly, arises from itinerant electronic states. The origin of the unusual Ising behavior  is analyzed as due to the peculiar near-Fermi edge nested electronic structure of {\ur} involving its strong spin-orbit interaction. The 
Ising anisotropy has pertinent implications for theories applicable to explaining the Hidden Order phase in {\ur}.

\end{abstract}
\maketitle

%
The origin of the ``Hidden Order" (HO) phase emerging below $T_0 = 17.5$~K in the uranium-based heavy-fermion compound {\ur} has remained a mystery even after more than a quarter century of intensive investigations (see, e.g.,  Ref.\ \cite{MydoshRMP11} for a recent review).
This second-order 
phase transition appears unmistakably in the thermodynamic and transport properties \cite{Palstra85, Schlabitz86, Maple86}, yet local solid-state probes such as x rays, neutron scattering, NMR or $\mu$SR fail to give a clue for the emerging order parameter. Long-range ordered (dipolar) magnetism has been excluded as a cause for the ``hidden order" (HO), but in close proximity to the HO phase a long-range ordered antiferromagnetic phase exists, which is stabilized through only a small pressure of $\sim 0.5$ GPa \cite{Amitsuka07}. 

Multifarious theories have been proposed to explain the intriguing appearance of the HO phase, see \cite{MydoshRMP11} and \cite{Chandra13} for an overview. 
Since the actinide $5f$ electrons can, in general, assume localized or itinerant character,  correspondingly theories adopting localized $5f$ behavior have been proposed (e.g., \cite{Chandra13,Ohkawa99,Haule09,Thalmeier11,Kusunose11,Hanzawa12}) as well as competing theories based on the assumption of  itinerant $5f$ behavior (e.g., \cite{Varma06, Elgazzar09, Dubi11,Fujimoto11,Pepin11,Ikeda12,Riseborough12,Das12,Rau12}). 
In several of the latter models the existence of a Fermi surface instability is typically connected to appearance of an unconventional density wave \cite{Riseborough12,Das12}, a spin resonance mode \cite{Elgazzar09} or hybridization wave \cite{Dubi11} that triggers formation of a Fermi surface gap. Theories based on localized $5f$ states often elaborate from a mainly localized $5f^2$ configuration  possibly with some hybridization with conduction electrons \cite{Chandra13,Hanzawa12}.

Experimental evidence in favor of either localized or itinerant $5f$ behavior is unquestionably crucial.
Recent quantum oscillation measurements \cite{altarawneh11,altarawneh12} have drawn attention to a previously unrecognized aspect of the HO quasiparticles in {\ur}, namely, their extreme Ising character. From the angular dependence of the de Haas-van Alphen amplitudes \cite{Ohkuni99} a $g$-factor anisotropy $g_c / g_a$, along the $c$ and $a$ crystallographic axes, was estimated to exceed 30, implying that HO emerges out of quasiparticles  with giant Ising anisotropy \cite{altarawneh11,altarawneh12}. This feature  has become salient
in the quest for understanding the exotic HO and its concomitant superconductivity 
\cite{Chandra13,Kusunose12}.
The Ising behavior of the near Fermi-energy quasiparticles nicely supports the picture of localized $5f$ states in {\ur}, possibly having a small hybridization with conduction electrons \cite{altarawneh11,altarawneh12}.
This extreme magnetic anisotropy is a central ingredient of the recent hastatic order theory in which a local 5$f^2$ crystal electrical field (CEF) doublet induces the Ising character \cite{Chandra13}. The Ising behavior might also be compatible with two singlet CEF states on the U$^{4+}$ ion that can 
form a hexadecapolar \cite{Haule09} or triakontadipolar order parameter \cite{Ressouche12},  but this was not yet shown.
For bandlike electrons, in contrast, a $g$-factor of 2 with little anisotropy would be expected  \cite{altarawneh12, Chandra13} which would render delocalized 5$f$ behavior unlikely.

Here we show on the basis of relativistic density-functional theory (DFT) calculations that the bandlike $5f$ electrons in {\ur} exhibit a colossal Ising behavior, a property which is truly exceptional for itinerant electrons. The origin of the unique Ising anisotropy is found to be due to a combination of the peculiar nesting of Fermi surface states and the strong spin-orbit interaction. Our results have important consequences for models applicable to unveil the nature of the HO. 

{\it Computational method} -- The DFT calculations were performed with the full-potential linearized augmented plane-wave (FP-LAPW) method as implemented in the {\sc wien}2k code (version v12.1) \cite{Blaha01} within the local-density approximation \cite{PW92}.
Spin-orbit (SO) coupling was self-consistently included with a second variational treatment \cite{kunes01b}. The employed atomic sphere radius $R_{MT}$ was 2.5 $a_0$ (Bohr radii) for U and Ru atoms, and 1.85 $a_0$ for Si. Our calculations were performed with a plane wave cut-off parameter $R_{MT}K_{_{max}}$ equal to 9.5, with $K_{_{max}}$ the maximum reciprocal space vector.   Relativistic local orbitals with a $p_{_{1/2}}$ radial wave functions were added to the uranium $6p$ semicore states. The total energy was converged to better than $1 \!\times \! 10^{-8}$ Ry, and the Brillouin zone is divided in $19\! \times \! 19 \! \times \! 8$ $k$ points. 
The crystallographic phase of {\ur} has, in the normal state above $T_0$, the body centered tetragonal structure; however, as has been emphasized in recent studies, the body-centered translation is broken in the HO phase \cite{Meng13} rendering the  unit cell (u.c.)\ similar to that of the antiferromagnetic phase, i.e., simple tetragonal (P4/mmm) with two inequivalent uranium atoms. Our calculations have been performed for the antiferromagnetic (AFM) phase for which it was recently clarified that its Fermi surface  (FS) is practically identical to that of the HO phase \cite{Meng13,Hassinger10,Oppeneer10}.

\begin{figure}[t!]
\begin{flushright}
\includegraphics[width=0.87\columnwidth,angle=0]{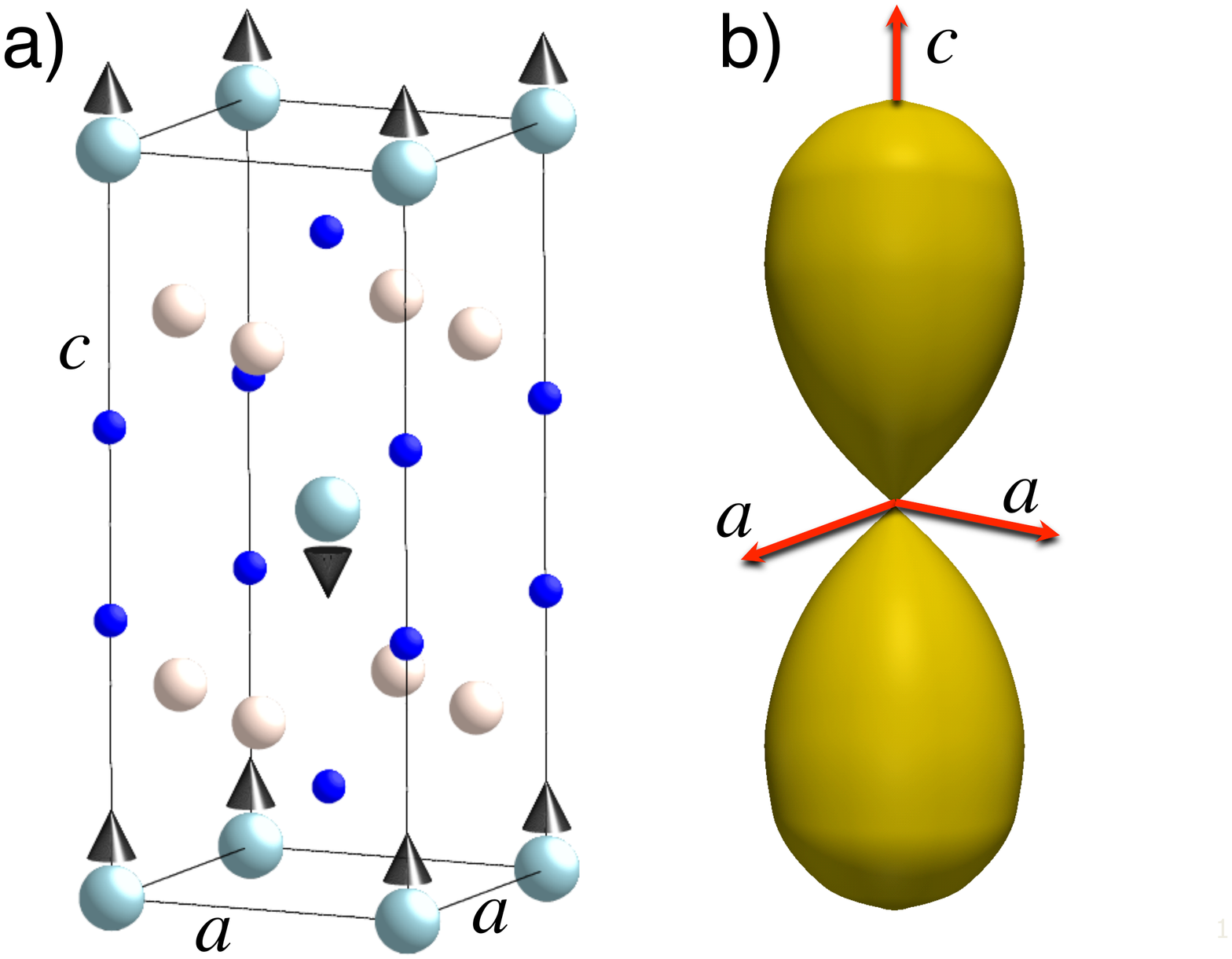}
\end{flushright}
\vspace*{-0.6cm}
\begin{center}
\includegraphics[width=0.95\columnwidth,angle=0]{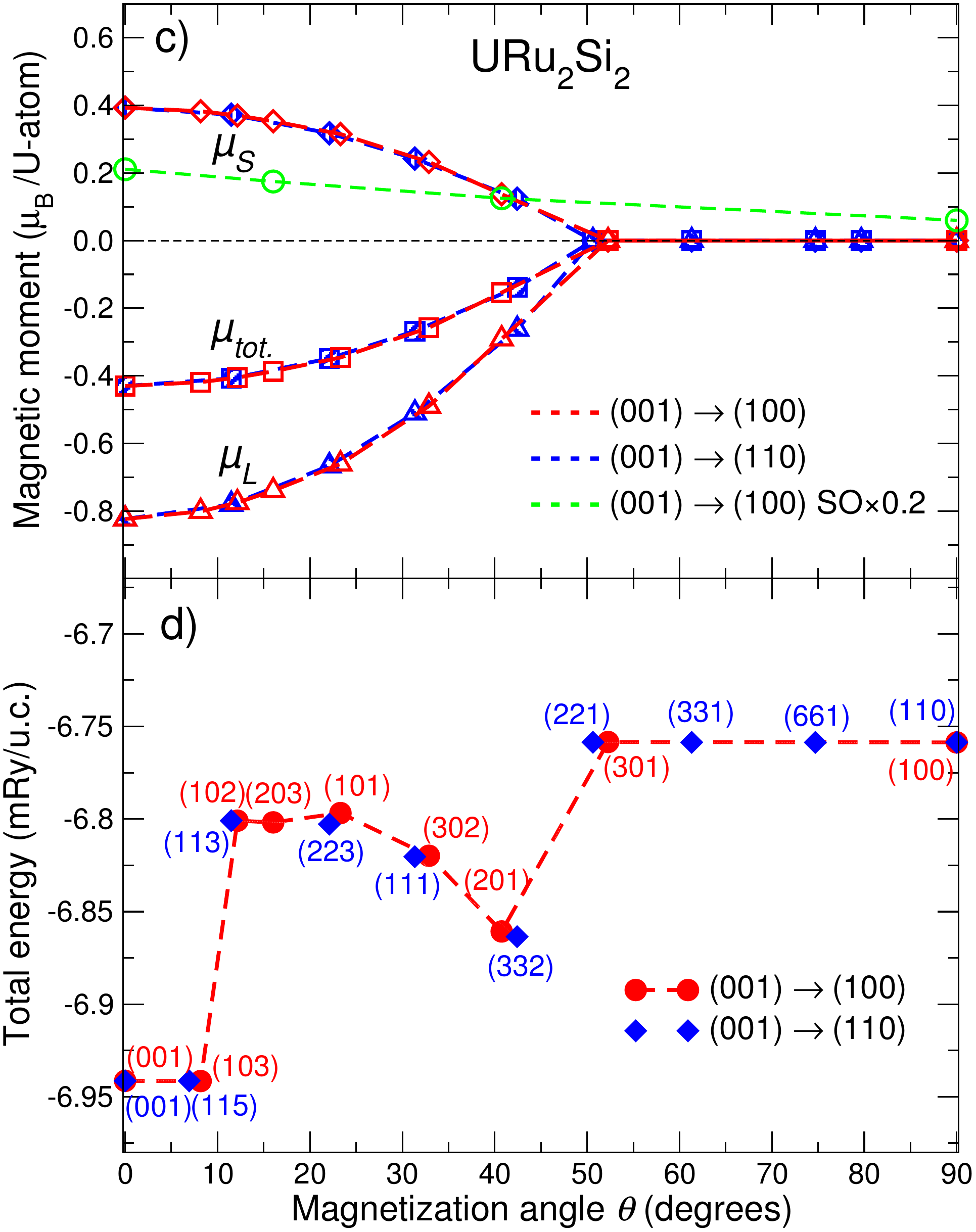}
\end{center}
\vspace*{-0.5cm}
\caption{(Color online) {\bf a)} The employed simple tetragonal unit cell of {\ur} with equilibrium directions of  uranium moments indicated. Large spheres depict U atoms, medium large spheres Ru, and small spheres Si atoms. {\bf b)} Three-dimensional plot of the magnitude of the total U moment calculated as function of the angles $\theta$ and $\phi$. \textbf{c)} Calculated dependence of the spin moment ($\mu_S$), total moment ($\mu_{tot.}$), and orbital moment ($\mu_L$) on the polar angle $\theta$, for two rotation directions in the unit cell. Also shown is $\mu_S$ calculated for a five-times reduced SO coupling.
{\bf d)} Computed total energy as a function of angle $\theta$ for two rotation directions, $(001)\rightarrow (100)$ and $(001) \rightarrow (110)$. The label at each symbol denotes the magnetization axis direction in Cartesian coordinates.}
\label{Fig1}
\end{figure}

{\it Calculated anisotropy} -- The anisotropy of the magnetic moment was computed by rotating the quantization axis stepwise from being parallel to the $c=(001)$ axis to lying in the tetragonal basal plane. See Fig.\ \ref{Fig1}(a) for a sketch of the tetragonal u.c.\ with equilibrium directions of moments indicated. At every axis direction--defined by the polar angle $\theta$ and azimuthal angle $\phi$--the electronic structure was computed self-consistently. The computed angular dependence of the total magnetic moment $\mu_{tot.}$ on one uranium atom  is shown in Fig.\ \ref{Fig1}(b). The magnitude of the moment is exceptionally anisotropic; the maximal total moment of 0.42\,$\mu_B$ is obtained when the magnetic axis is along the tetragonal $c$ axis, but the moment vanishes for directions approaching the basal plane. In contrast to the marked dependency on the polar angle, the magnetization does not show a notable variation with the azimuthal angle $\phi$. The calculated dependence of the total moment, and spin ($\mu_S$) and orbital ($\mu_L$) angular moments 
on $\theta$ is plotted in Fig.\ \ref{Fig1}(c), for two rotation directions in the unit cell, $(001)\rightarrow (100)$ and $(001) \rightarrow (110)$. Note that the orbital moment is opposite to the spin moment and twice larger, which emphasizes the importance of accounting for the strong SO coupling of uranium in this material. The moments continuously decrease with $\theta$: a tilt of the moment by 35$^{\circ}$ off the $c$ axis reduces it by 50$\%$ and it completely collapses at $\theta =50^{\circ}$, for both $\phi$ angles.
This highlights the extreme Ising anisotropy calculated here for URu$_2$Si$_2$. A similar uniaxial Ising anisotropy has never been previously reported for any material.

The small uranium moment stipulates that the here-appearing magnetism is band magnetism, in contrast to the large atomiclike moment due to on-site Coulomb interaction that is e.g.\ found for UO$_2$. For band magnetism the  long-range  exchange interaction is important, which typically shows an oscillatory behavior due to the FS. 

Figure \ref{Fig1}(d) gives the computed total energy as a function of angle $\theta$, for two directions in the u.c. Again we observe that there is practically no dependence on $\phi$, but the total energy does not vary smoothly with the angle $\theta$ as the moments do. The minimum of the total energy is confined to a narrow region of $\theta \le 10^{\circ}$, i.e., for the magnetization axis nearly along the $c$ axis. An intermediate minimum occurs around $40^{\circ}$. The total energy increases somewhat abruptly beyond 10$^{\circ}$ and $47^{\circ}$, after which it stays constant. The latter increase of the total energy occurs at the same angle where the moment vanishes. 

{\it Analysis} -- Before examining the origin of the uniaxial Ising anisotropy it is important to realize that such behavior is exceptional. Particularly, in spite of extensive theoretical studies of the Ising model, there are few three-dimensional (3D) Ising materials.  Ising anisotropy is known to occur in 1D metalorganic compounds (e.g., \cite{Caneschi01}) and also in transition metal oxides containing chain structures \cite{Coldea10},  but there is not a single 3D metallic material known that exhibits such an extreme Ising anisotropy as computed here for {\ur}.  For comparison, the magnetically most anisotropic material with itinerant electrons is presently FePt \cite{Weller99} which moreover, crystallizes in the simple tetragonal structure too, with equilibrium moments along $c$. However, the calculated Fe total moment is hardly anisotropic (2.863 $\mu_B$ vs.\ 2.857 $\mu_B$ for $\mu$ along (001) and (110), respectively, see \cite{Oppeneer98}), stipulating that the magnetic behavior is more Heisenberg than Ising-like. Thus FePt is consistent with the observation in Refs.\  \cite{altarawneh12,Chandra13} that for an itinerant electron material an isotropic $g$-factor would be expected. 
The Ising anisotropy in {\ur} is furthermore unusual because it is obtained here for an AFM alignment of U moments. For {\ur} with inversion symmetry this implies that the band dispersions are four-fold Kramers degenerate. The band degeneracy is not lifted by rotation of the moment, in contrast to SO-related degeneracies in ferromagnetic materials. 

\begin{figure}[tbp]
\begin{center}
\includegraphics[width=0.95\columnwidth,angle=0]{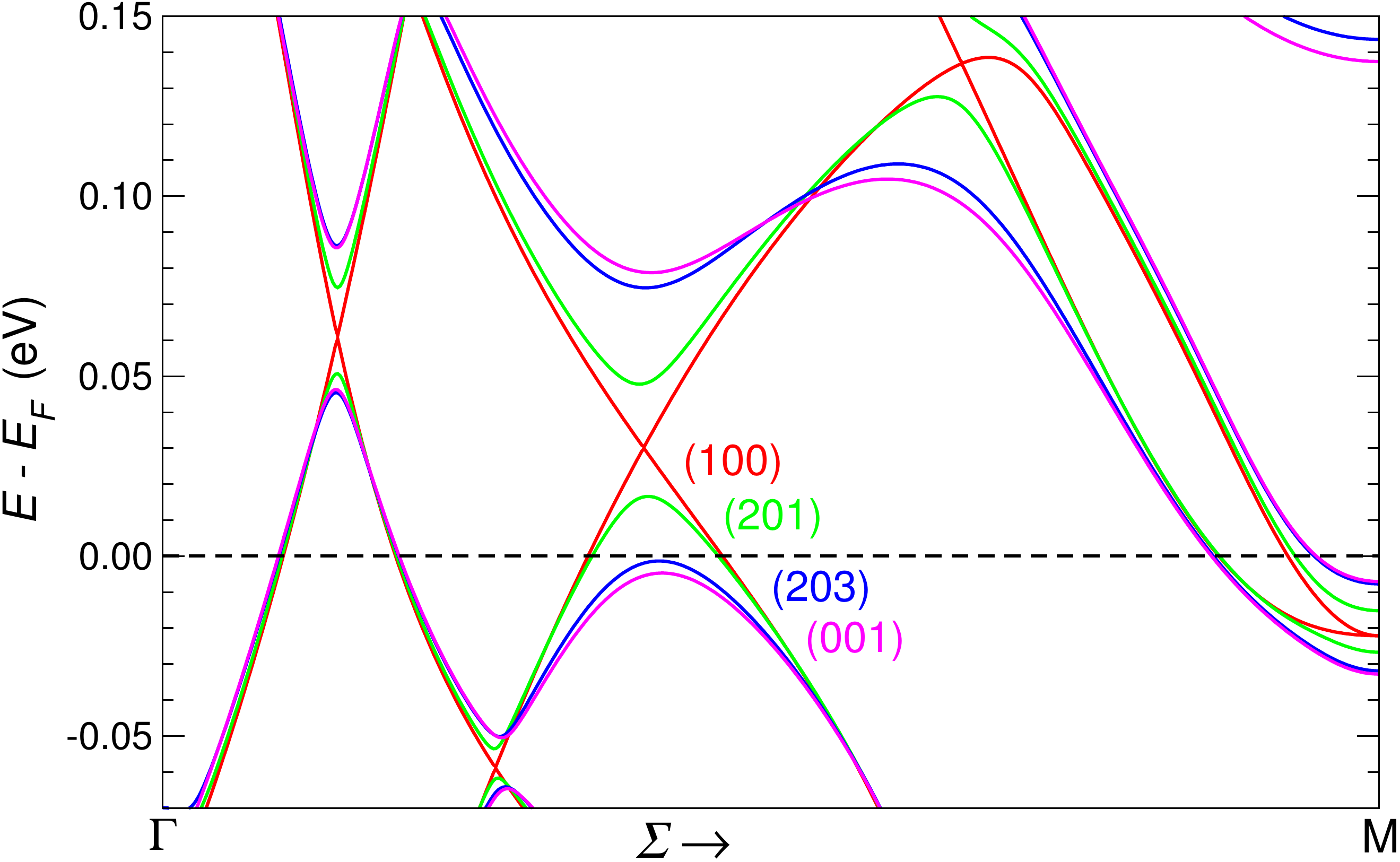}
\end{center}
\vspace*{-0.5cm}
\caption{(Color online) Computed dependency of the energy band dispersions of URu$_2$Si$_2$ on the polar angle $\theta$. Shown are the dispersions along the $\Sigma$ ($\Gamma \rightarrow$\,M) high-symmetry direction in the simple tetragonal Brillouin zone. The labels give the directions of the magnetic axis in Cartesian coordinates.}
\label{Fig2}
\end{figure}

To analyze the origin of the Ising behavior we consider first the SO interaction. To show its effect we  artificially reduced this term in the calculations. The results of a calculation with a five-times reduced SO interaction are given in Fig.~1(c). The anisotropy of $\mu_S$ becomes strongly reduced. Upon reducing the SO interaction to zero we obtain $\mu_S (100)$=$\mu_S (001)$=0.105\,$\mu_B$, i.e., the moment has become entirely isotropic. 
This emphasizes that the strong SO interaction of U is indispensable for the Ising anisotropy.
Next,
we plot in Fig.\ \ref{Fig2} the band dispersions near the Fermi energy ($E_F$) along the $\Sigma$ high-symmetry direction in the Brillouin zone. It is along this direction that the previously reported FS gapping appears \cite{Elgazzar09,Oppeneer10}.
A peculiar nesting situation of two bands which have almost pure uranium $j_z$=$\pm 5/2$ and $\pm3/2$ character leads to a protected Dirac crossing point which, in the nonmagnetic state, falls just above $E_F$ along the $\Gamma -$\,M direction. However, at several low-symmetry directions it lies precisely on $E_F$ \cite{Oppeneer10,Oppeneer11}.
When the moments are oriented along (001) the Dirac crossing is maximally lifted and the lower $j_z$=$\pm 3/2,\,\pm 5/2$ hybridized band falls below $E_F$. Upon rotating the magnetic axis to (203) the lower hybridized band shifts closer to $E_F$ and the gap along the $\Sigma$ direction is reduced, see Fig.\ \ref{Fig2}. As will become clear below,  this band already crosses $E_F$ at other parts in the Brillouin zone.
Rotating the magnetic axis further to (201) destroys the FS gapping along the $\Sigma$ high-symmetry direction, but there is still a lifting of the degeneracy. This small degeneracy lifting disappears when the quantization axis is rotated further to (100), where the dispersions become equal to those of the normal nonmagnetic state.

\begin{figure}[tb]
\includegraphics[width=1.0\columnwidth,angle=0]{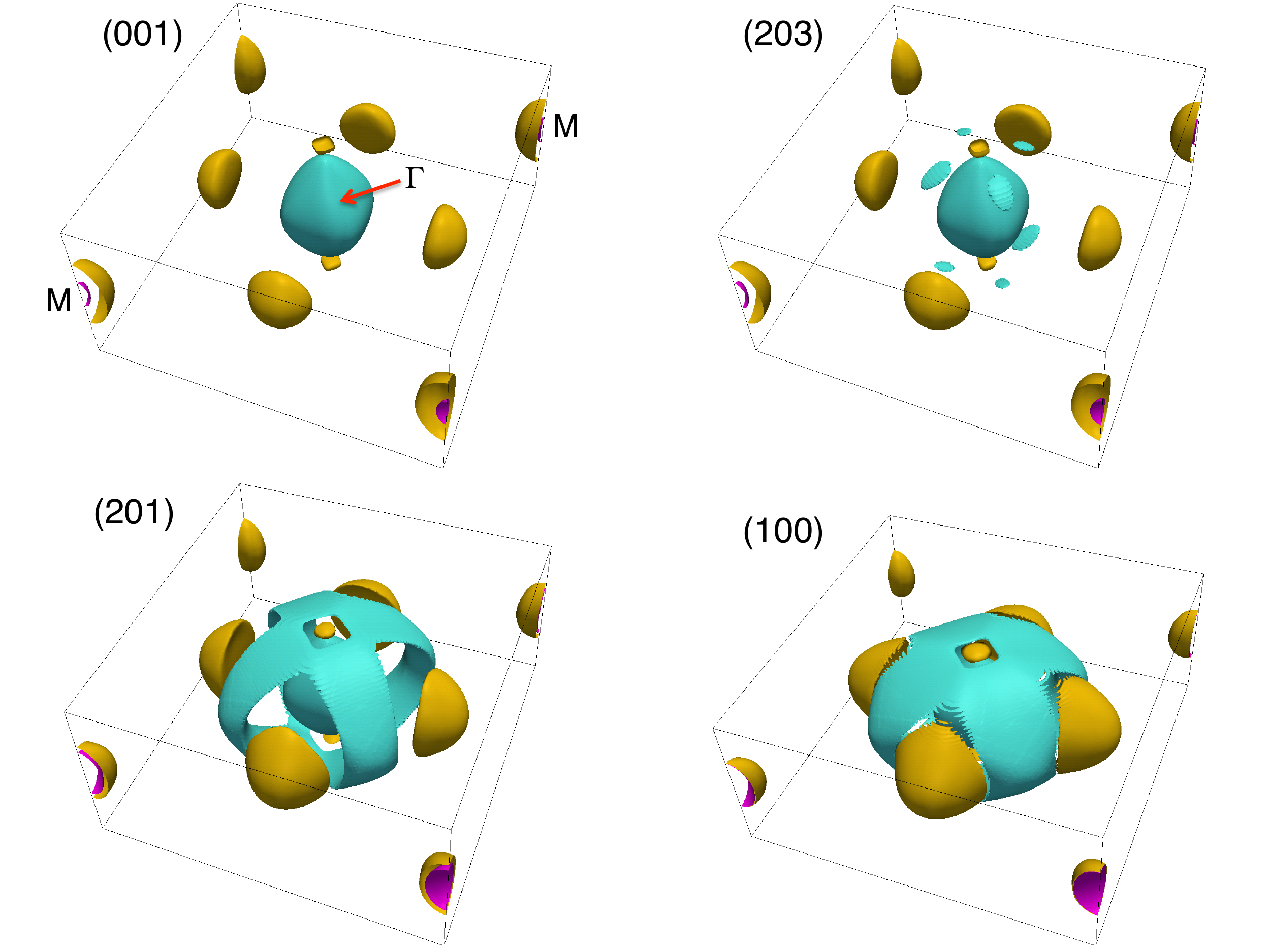}
\vspace*{-0.5cm}
\caption{(Color online) Calculated Fermi surfaces for various representative orientations of the magnetic axis, denoted by the top-left labels. The Fermi sheets due to different bands are depicted by the three different colors. Under optimal nesting condition for moment along (001) the light green, cagelike Fermi sheet is completely gapped. Turning the magnetization axis to (203) and (201) destroys the nesting, until the gapping has practically disappeared for an in-plane moment (100), where the cagelike Fermi surface completely appears.}
\label{Fig3}
\end{figure}

Further information on the dependence of the electronic structure on the magnetic axis can be obtained from the calculated FSs, shown in Fig.\ \ref{Fig3}. The FS of {\ur} in the \textsc{st} Brillouin zone consists of two electron pockets at the M point, a larger $\Gamma$-centered hole pocket, with a small $\Gamma$-centered pocket inside of it,  and four hemispherical electron pockets \cite{Elgazzar09}. This FS  topology  is supported by quantum oscillation and 
ARPES measurements \cite{Hassinger10,Meng13}.
There exists in addition  a cagelike FS sheet which only appears for certain magnetization axes. The cagelike FS sheet and the four hemispherical pockets result from two bands with $j_z$=$\pm 5/2$ and $\pm 3/2$ character in the normal state which exhibit particular nesting properties \cite{Oppeneer11}.  For $\mu\,||\,(001)$ 
the cagelike FS sheet is fully gapped, which is consistent with recent ARPES that did not detect this sheet in the HO phase \cite{Meng13}. Upon rotation of the magnetic axis the full gapping is destroyed
and small patches of the cage FS sheet appear for $\mu || (102)$ or larger, which corresponds to the steplike increase of the total energy in Fig.\ \ref{Fig1}(d). The area of the cage FS sheet increases further with rotation of the quantization axis until direction (201), where the gapped FS area has become small. For larger polar angles the FS collapses to that of the high-temperature nonmagnetic state \cite{note}.
Note that the FS behavior of {\ur} at the HO transition is \textit{converse} to the standard behavior expected in Kondo lattices, where the small FS is expected at elevated temperature and the large FS at low temperature.
 
 {\it Discussion} -- Our calculations show that {\ur} exhibits a unique 3D Ising anisotropy which is unusual for bandlike electrons. {\ur} is however special, first, because the SO splitting of the $5f$ states is about 0.8 eV, whereas their exchange splitting is only about 0.1 eV.  Thus, due to the uranium SO interaction the electronic structure couples significantly to the quantization axis. This is different from the aforementioned FePt, where the exchange splitting is much larger than the SO splitting. Second, the peculiar, strongly nested near-Fermi edge electronic structure provokes the Ising anisotropy. Importantly, since the HO and AFM phases of {\ur} share the same FS and SO interaction, the observed Ising behavior can be extended to the HO phase.
 
 The computed Ising anisotropy tallies well with the $g$-factor anisotropy deduced recently  from quantum oscillation measurements \cite{altarawneh12}. Here a polar plot of the $g$-factor anisotropy resulted in a figure ``8" shape, which reasonably compares to the dumbbell-shaped moment anisotropy in Fig.\ \ref{Fig1}(b). The latter shape is narrower (i.e., more Isinglike),
which can be due to the fact that a different quantity is studied ($\mu$ vs.\ $g$-factor). The angular dependence of the spin magnetic moment  $\mu_S$=$\chi_{_S} H$ can be estimated from the dependence of the spin susceptibility on the polar angle $\theta$, which is given 
in Ref.\ \cite{altarawneh12} 
(notably only for one quantum oscillation orbit) as $\chi_{_S}$$\propto g_c^2 \cos^2 \theta + g_a^2 \sin^2 \theta$, with $g_c$=2.65 and $g_a$$\approx$0.0. The computed total moment in Fig.\ \ref{Fig1}(c) varies as
 $\mu_{tot.} (\theta)\approx$ $\mu_c \cos 2\theta =$$\mu_c (\cos^2 \theta - \sin^2 \theta)$ for $\theta \le 45^{\circ}$, 
 having thus the same leading term $\cos^2 \theta$ for moderately small $\theta$.  
 
Isinglike behavior has also been observed  in other properties of {\ur} \cite{MydoshRMP11}.
Neutron scattering revealed that magnetic resonance modes in the HO phase are both itinerant and strongly Isinglike \cite{Wiebe07,Bourdarot11}. The gapping of itinerant spin excitations was shown \cite{Wiebe07} to account completely for the entropy loss at the HO transition \cite{Maple86}. The appearance of such excitations is compatible with the here-computed electronic structure; the FS sheets are nested and each one is typified mainly by one kind of U $j_z$ character. The resonance mode at $\boldsymbol{Q}_0$=(0,\,0,\,1) could be assigned to Isinglike spin-orbital excitations between FS sheets with $j_z$=$\pm$5/2 and $\pm$3/2 character, and the resonance at
$\boldsymbol{Q}_1$=(1.4,\,0,\,0) to sheets with $j_z$=$\pm$3/2 and $\pm$1/2 character \cite{Oppeneer11,Das12}.

An important dichotomy in the on-going debate on the origin of the HO is, whether the uranium $5f$  electrons are localized or  itinerant.
The Ising anisotropy of quasiparticles has recently gained considerable weight in this discussion. 
 It was emphasized that this Ising anisotropy is 
  a fingerprint of a localized $5f^2$ non-Kramers 
 doublet whose corresponding local-moment anisotropy in the crystal field was demonstrated to imprint a comparable $g$-factor anisotropy  \cite{Chandra13}. These results thus strongly advocated the picture of localized $f$ electrons in {\ur}. Several other theories (e.g., \cite{Santini94,Haule09,Kusunose11,Ressouche12}) are based on other choices of the CEF levels, thereby leading to a variety of multipolar orders 
 proposed to explain the HO.
However, our study proves that the extreme Ising anisotropy can arise equally well from itinerant electrons.  

Since both the itinerant and hastatic localized model can explain this feature, further experimental arguments need to be brought to bear on the debate.
 Recent analyses of available data clarified that many properties of {\ur} are compatible with the picture of itinerant $f$ electrons \cite{Meng13,Oppeneer11}, while CEF excitations characteristic of localized $f$ electrons could not be detected \cite{Wiebe07}.
Furthermore, recent resonant x-ray \cite{Walker11} and neutron scattering 
\cite{Khalyavin14}  experiments could not confirm the presence of quadrupolar, octupolar or triakontadipolar ordering, and neither could the in-plane moment predicted for hastatic order \cite{Chandra13} be detected
\cite{Das13}.
 Our results hence underline that 
the itinerant picture is the suitable starting point for explanations  of the HO, which is best viewed as a FS reconstruction emerging out of delocalized $5f$ states.
 
To conclude, our study reveals that {\ur} is an exceptional material in which a giant 3D Ising anisotropy arises from bandlike electronic states. The
Ising character and the HO phase are two unique features of {\ur} and the question naturally emerges how, and if, they are related. The Ising nature moreover puts a rigorous constraint on theoretical proposals for the HO phase, as any relevant theory must account for this unusual feature. 
 


We thank A.\ Aperis and B.\ A.\ Ivanov for 
discussions.
This work was supported through the Swedish Research Council (VR), the G.\ Gustafsson Foundation, and the Swedish National Infrastructure for Computing (SNIC).
{}


\begin{thebibliography}{}
\bibitem{note}
Due to the finite number of $k$ points used to plot the FS the light green and golden FS sheets do not fully join.
\end{thebibliography}
\end{document}